\documentstyle[12pt]{article}

\font\tenrm=cmr10

\font\elevenbf=cmbx10 scaled\magstep 1
 1
\font\elevenit=cmti10 scaled\magstep 1

\begin{document}
\thispagestyle{empty}
\hfill{CERN-TH/95-103}
\vskip 0.1cm
\hfill{SMU HEP 95-02}
\vskip 0.5cm
\begin{center}{{CHARM PRODUCTION IN DEEP-INELASTIC $e\gamma$
SCATTERING\\ TO NEXT-TO-LEADING ORDER IN QCD}
\footnote{Talk presented by E. Laenen at Photon'95, Sheffield, UK,
April 8-13, 1995.}
}
\vglue .5cm
\vglue 0.8cm
\begin{sc}
 Eric Laenen\\
\end{sc}
{\sl CERN TH-Division\\
1211-CH, Geneve 23, Switzerland}
\vglue 0.15cm
and\\
\vglue 0.2cm
\begin{sc}
Stephan Riemersma\\
\end{sc}
{\sl
Department of Physics, Fondren Science Building\\
Southern Methodist University, Dallas, TX 75275, U.S.A}
\vglue 0.5cm
\end{center}

\vglue 1cm
\begin{abstract}
\par \vskip .1in \noindent
We discuss the calculation of $F_2^{\gamma}({\rm charm})$
to next-to-leading order (NLO) in QCD, including contributions
from both hadronlike and pointlike photons. We show that the former
dominates strongly below $x\simeq 0.01$, and 
the latter above this value.
This fact makes $F_2^{\gamma}({\rm charm})$ for 
$x \geq 0.01$ calculable, whereas for 
$x \leq 0.01$ it serves to constrain the small-$x$ 
gluon density in the photon. Both ranges in $x$
are accessible at LEP2.
Theoretical uncertainties are well under
control. We present rates for single-tag
events for the process 
for $e^+e^- \rightarrow e^+e^- c X$ for LEP2.
Although these event rates are small, we 
believe a measurement of $F_2^{\gamma}({\rm charm})$
is feasible. 
\end{abstract}

\vfill{CERN-TH/95-103
\vskip 0.1cm
SMU HEP 95-02
\vskip 0.1cm
May 1995}

\newpage
\setcounter{page}{1}

\noindent Open heavy quark production in two-photon collisions 
at $e^+e^-$ colliders has been
difficult to observe in experiments. This is mostly
due to the typically small cross section 
coupled with
low charm tagging efficiencies.
At LEP1 an additional difficulty is represented by the
$Z^0$ background, from which a two-photon sample 
is hard to isolate.
Experimentally one has studied the reaction
$e^+e^- \rightarrow e^+e^- D^{*\pm} X$ with neither outgoing
lepton tagged (``no-tag''), because it
proceeds predominantly via the fusion of two equivalent photons 
to produce open charm ($\gamma\gamma\rightarrow c\bar c$).
Measurements of the $D^{*\pm}$ cross section
in no-tag  $e^+e^-$ collisions have been performed by
JADE, TASSO, TPC/2$\gamma$, TOPAZ, VENUS, ALEPH
and AMY.
\cite{cX}. 

The equivalent measurement for the case in which
one of the outgoing leptons is tagged (``single-tag'')
has not been performed.
The demand for the extra tag 
suppresses the rate too much for present experiments to measure.

The theory is in better shape. The cross section
for $\gamma\gamma\rightarrow c\bar c$ has
been calculated to next-to-leading (NLO) order in QCD
in \cite{DKZZ}, and found to agree quite well with experimental
results. Here we focus on the single-tag case.
We begin in general by considering the reaction
\begin{equation}
e^-(p_e) + e^+ \rightarrow e^-(p_e') + e^+ + X\,,\label{one}
\end{equation}
where $X$ denotes any hadronic state 
allowed by quantum-number 
conservation laws.
When the outgoing electron is tagged then
this reaction is dominated by the subprocess
\begin{eqnarray}
\gamma^*(q) + \gamma(k) \rightarrow X\,,\label{two}
\end{eqnarray}
where one of the photons is highly virtual and the other one
is almost on-mass-shell. The case where the positron is 
tagged is completely equivalent. This process is described by the
cross section
\begin{eqnarray}
\frac{d^2\sigma}{dxdQ^2} &=& \int dz \, z\, f_\gamma^e (z, \frac{S}{m_e^2})
\,  \frac{2\pi\alpha^2}{x\,Q^4} \nonumber \\&&
\times \left[ \{ 1 + (1-y)^2\}
F^\gamma_2(x,Q^2) -y^2 F^\gamma_L(x,Q^2) \right]\,,\label{three}
\end{eqnarray}
where the $F^\gamma_k(x,Q^2)$ $(k=2,L)$ denote the deep-inelastic
photon structure functions and $\alpha = e^2/4\pi$
is the fine structure constant.
The Bjorken scaling variables $x$ and $y$ are
defined by
\begin{equation}
x = \frac{Q^2}{2k\cdot q} \,, \qquad y = \frac{k\cdot q}{k \cdot p_e}
\,, \qquad q = p_e - p_e'\,, \label{four}
\end{equation}
where $p_e$, $p_e'$ are defined in (\ref{one}).
The off-mass-shell photon and the on-mass-shell photon
have four-momenta $q$ and $k$ 
respectively with $q^2 = -Q^2 <0$
and $k^2 \approx 0$. Because the photon with momentum $k$ is 
almost on-mass-shell, eq.~(\ref{three}) is written in the Weizs\"acker-Williams
approximation: the function $f^e_\gamma(z,S/m_e^2)$ represents
the probability of finding a photon $\gamma(k)$ in the positron,
with longitudinal momentum fraction $z$, and is given in first approximation
by
\begin{equation}
f^e_\gamma(z,\frac{S}{m_e^2}) = \frac{\alpha}{2\pi}
\frac{1 + (1-z)^2}{z} \ln \frac{(1-z)(zS-4m^2)}{z^2m_e^2}\,,\label{five}
\end{equation}
provided a heavy quark with mass $m$ is produced
(for light quarks $m=0$).
Here $S$ is the c.m. energy squared of the electron-positron system.

In (\ref{three}) both structure functions can be represented as
\begin{equation}
F_k^{\gamma}(x,Q^2) = \sum_{i=q,\bar q, g} f_i^{\gamma}
\otimes C_{k,i}(\frac{Q^2}{\mu_f^2}) +
C_{k,\gamma}(\frac{Q^2}{\mu_f^2}). \label{six}
\end{equation}
The $f_i^{\gamma}$ are photonic parton densities and
the $C_{k,i}\,(i=q,\bar q,g,\gamma)$ are Wilson coefficient functions.
For the results presented here we used for the $f_i^{\gamma}$ 
the GRV leading-order (LO) set for LO calculations, and the GRV
higher-order (HO)
set for NLO ones \cite{GRV}, in the $\overline{\rm MS}$ scheme.
In \cite{LRSN} 
$F_2^{\gamma}$ and $F_L^{\gamma}$ were calculated to 
NLO in QCD by computing all $O(\alpha_s)$ corrections
to the coefficient functions $C_{k,i}$. 
For $F_L^{\gamma}$ this represented the first NLO
analysis, while for $F_2^{\gamma}$ the new
corrections consisted of those due to the inclusion 
of heavy quarks. We will now focus on 
these heavy quark contributions to $F_2^{\gamma}$ for the case
of charm, and highlight some
of the very interesting features they possess.

Note that in (\ref{six}) both terms on the right hand side
depend on the mass factorization scale $\mu_f$, which,
although it may be judiciously chosen, is in principle
arbitrary. If one now demands the presence of a
charm quark in the final state, the $\mu_f$ dependence
of the second term is not present through NLO. Splitting
$F_2^{\gamma}$ then according to (\ref{six}) as
\begin{equation}
  F_2^{\gamma}({\rm charm}) = F_2^{\gamma,HAD}(x,\frac{Q^2}{\mu_f^2}, 
\frac{m^2}{\mu_f^2})
+ F_2^{\gamma,PL}(x,\frac{Q^2}{m^2}) 
\end{equation}
this means that $F_2^{\gamma,PL}$ is completely
calculable, whereas $F_2^{\gamma,HAD}$ is, 
analogously to the proton case, mainly sensitive to the
photonic gluon density.

\vglue 6.5cm

\vbox{\includegraphics{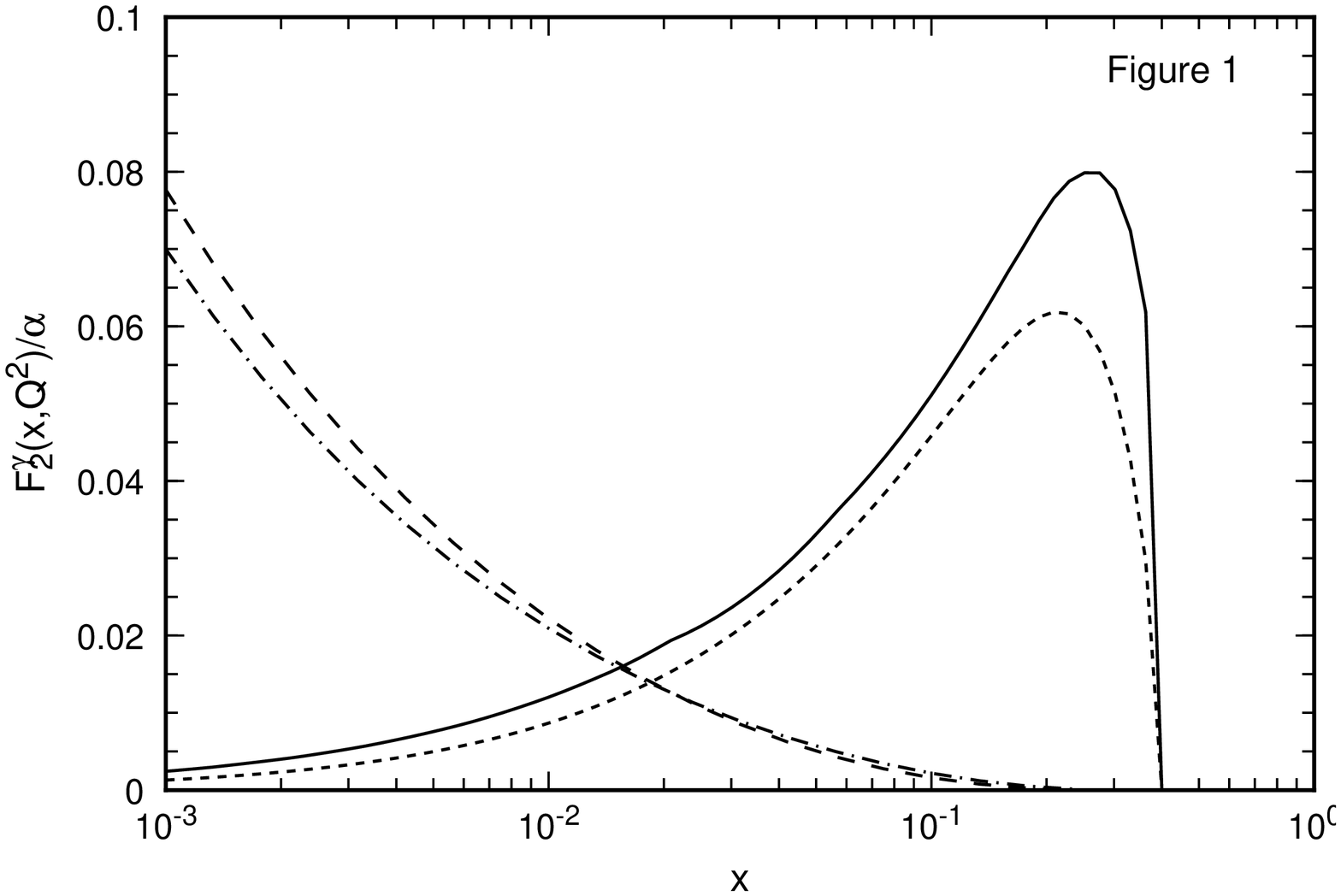}}      

\vglue 6.1cm

\vbox{\includegraphics{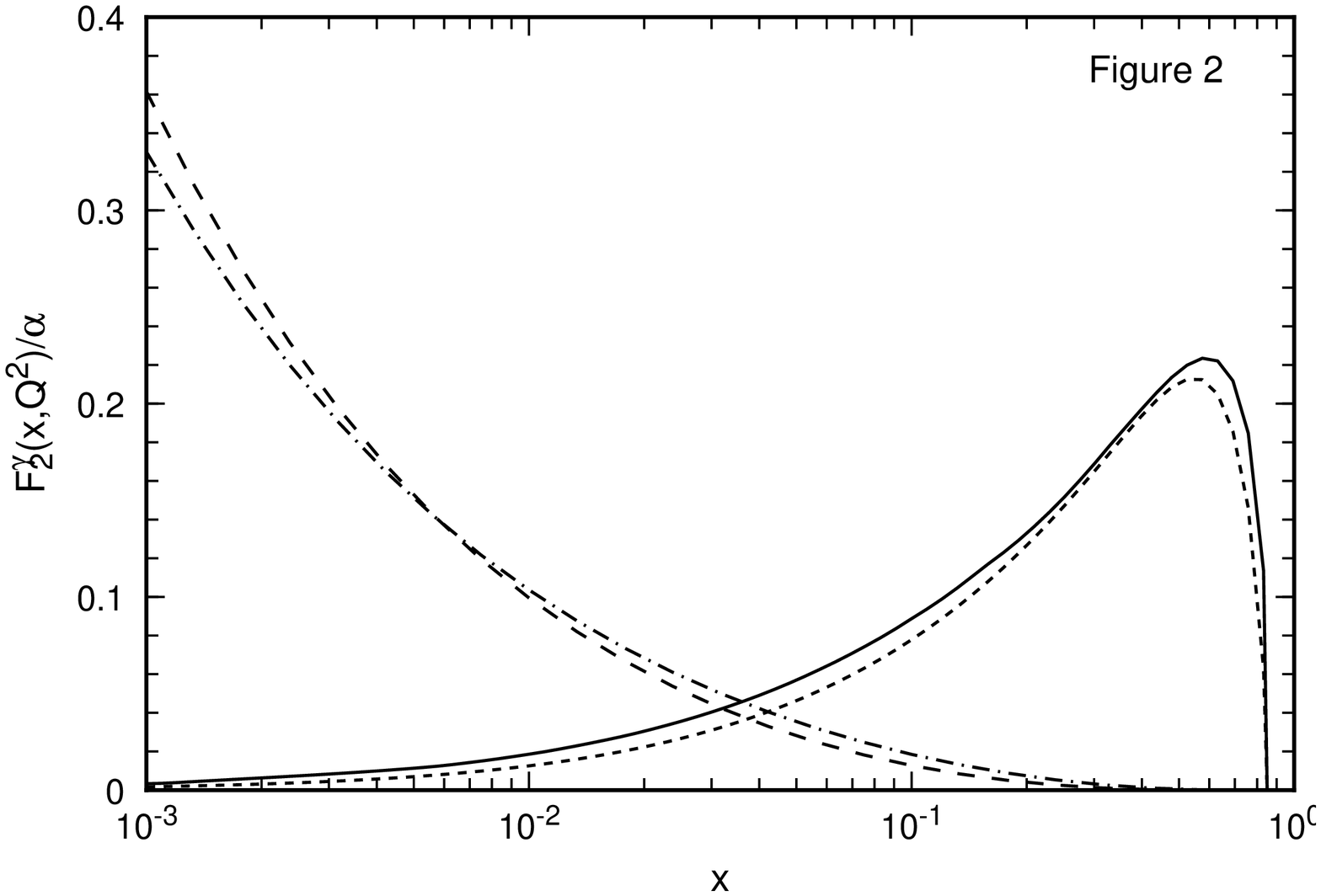}}      

{\tenrm\baselineskip=12pt 
\noindent Figure 1: Hadronic and pointlike contributions
to $F_2^{\gamma}({\rm charm})/\alpha$ at $Q^2= 5.9\,({\rm GeV}/c)^2$.
The short-dashed line denotes the LO pointlike, and the solid line the
NLO pointlike component. 
The long-dashed line denotes the LO hadronlike, and the dot-dashed the
NLO hadronlike component. 
Figure 2: same as in Figure 1, now at $Q^2= 51\,({\rm GeV}/c)^2$.
For both figures we used a one-loop running $\alpha_s$ with $\Lambda=
232\,{\rm MeV}$ at LO and a two-loop running  $\alpha_s$ with $\Lambda=
248\,{\rm MeV}$ at NLO, both for three active flavors.
We used a charm mass of $1.5\,{\rm GeV}$.}

\vglue .5cm
The only uncertainties in $F_2^{\gamma,PL}(x,Q^2)$ are
due to $\alpha_s$ and the charm mass.
At a value of $Q^2 \approx 
6\,({\rm GeV}/c)^2$ the uncertainty due to the latter
is about 20\%, and about
10\% for $Q^2 \approx 50\,({\rm GeV}/c)^2$, if the 
charm mass is varied between 1.3 and 1.8 ${\rm GeV}$.
The largest uncertainty in $F_2^{\gamma,HAD}$ is
due to the small-$x$ photonic gluon density $f_g^{\gamma}$,
hence a measurement of  $F_2^{\gamma,HAD}$ could be used to 
constrain $f_g^{\gamma}$. 

Figs. 1 and 2 show $F_2^{\gamma,PL}$ and $F_2^{\gamma,HAD}$
versus $x$ at both LO and NLO, for $Q^2= 5.9\,({\rm GeV}/c)^2$
and $Q^2= 51\,({\rm GeV}/c)^2$ respectively. Note that
the $O(\alpha_s)$ corrections to $F_2^{\gamma,PL}$ are 
fairly small.
We may thus assume that even
higher order contributions are negligible,
and that thus this component
is calculated with some precision.

The most remarkable feature in these figures is clearly
the clean separation in $x$ of both components. As a result
on may confront a precise calculation and constrain
the small-$x$ photonic gluon density in one experiment.

As mentioned earlier, such an experimental study is very difficult 
in practice, due
to the inherently low event rate and the difficulty of
efficient charm tagging. In order to judge
the feasibility of such a study we have integrated
(\ref{three}) for various $x,Q^2$ bins, and obtained
estimates for the number of charm quarks per bin produced
at LEP2.

As an aside, in order to perform the integrals over $x$ and $Q^2$,
we used fitted versions of the coefficient functions
$C_{k,i}$ in (\ref{six}), as the actual expressions
in \cite{LRSN} are too long for fast evaluation.
By adapting the fitted coefficient functions of
electroproduction of heavy quarks on a proton
target \cite{RSN} to our case we were able to speed up the code
by as much as a factor of twenty. 

For the Weizs\"acker-Williams density of 
equivalent photons in the electron
we used here the improved
version of \cite{FMNR}, which allows for an anti-tagging angle.
Furthermore, in order to test the stability of the results
we put the renormalization scale equal to the mass factorization
scale $\mu_f$ and varied the latter 
from $Q/2$ to $2Q$.

The results are shown in Table 1. 
We have shown here the total number of events in the
top half of the table, and its
pointlike component in the bottom half. The 
hadronlike component per bin is obtained by simply taking the difference.
We see that even with
a charm tagging efficiency of, say, 2\% on average a few tens of 
events per bin should be observable for larger $x$
values. The numbers in Table 1 increase by about 40\% 
if one uses the na\'ive Weizs\"{a}cker-Williams density of
(\ref{five}). 
Note that they change relatively little under variations in 
the factorization scale $\mu_f$.

Given the small event rates and the interesting physics
these events provide a handle to, 
it is clearly important to obtain the
highest possible charm tagging efficiency at LEP2.
The results of various methods of charm tagging 
have been presented at this conference \cite{tag} 
and there are prospects to achieve a satisfactory
efficiency at LEP2.

Summarizing, we have shown the clear separation in
$x$ of the contributions to $F_2^{\gamma}({\rm charm})$
due to pointlike and hadronlike photons.
The pointlike component is calculable
in perturbative QCD, while the hadronlike component constrains the 
small-$x$ gluon density in the photon.. An estimate of event
rates suggests that a measurement of $F_2^{\gamma}({\rm charm})$,
although difficult, may be feasible at LEP2, and is certainly
worthwhile.

\newpage

{\tenrm\baselineskip=12pt Table 1. Total number of events for single-tag
charm production at LEP2 ($500/pb$ integrated luminosity), based on NLO
QCD, determined from eq. (\ref{one}). We used the Weizs\"{a}cker-Williams density
of Frixione et al. \cite{FMNR} with an anti-tagging angle
$\theta_c = 20 \, {\rm mrad}$, and the GRV HO
set of photonic parton densities, in the $\overline{\rm MS}$ scheme
\cite{GRV}.}
\begin{center}
\begin{tabular}{||c|c|c|c|c|c||} \hline \hline
$Q^2$     &  $Q^2$       &  $x$                 & \multicolumn{3}{c||}{Events}  \\ \cline{4-6}
(GeV$^2$) &  range       &  range               & $\mu_f = Q/2$ & $\mu_f = Q$ & $\mu_f = 2Q$ \\ \hline
          &              &  Total               &               &             &              \\ \hline
 6.6      & 3.2 - 10     & $ 1.0 - 3.2 \cdot 10^{-4}$ & 47 & 45 & 44 \\ 
          &              & $ 3.2 - 10.0 \cdot 10^{-4}$ & 294 & 273 & 264 \\ 
          &              & $ 1.0 - 3.2 \cdot 10^{-3} $ & 571 & 517 & 494 \\ 
          &              & $ 3.2 - 10.0 \cdot 10^{-3}$  & 783 & 699 & 659 \\ 
          &              & $ 1.0 - 3.2 \cdot 10^{-2}$ & 1104 & 1002 & 949 \\ 
          &              & $ 3.2 - 10.0 \cdot 10^{-2}$  & 2091 & 1969  & 1904  \\ 
          &              & $ 1.0 - 3.2 \cdot 10^{-1}$  & 4403 & 4057  & 3873  \\ 
          &              & $ 3.2 - 10.0 \cdot 10^{-1}$  & 832 & 718 & 656 \\ \hline
 20.8     & 10 - 32      & $ 3.2 - 10.0 \cdot 10^{-4}$ & 24 & 24 & 24 \\ 
          &              & $ 1.0 - 3.2 \cdot 10^{-3}$ & 144 & 138 & 136 \\ 
          &              & $ 3.2 - 10.0 \cdot 10^{-3}$  & 274 & 260  & 253  \\ 
          &              & $ 1.0 - 3.2 \cdot 10^{-2}$  & 409 & 384  & 371  \\ 
          &              & $ 3.2 - 10.0 \cdot 10^{-2}$  & 713 & 683  & 666  \\ 
          &              & $ 1.0 - 3.2 \cdot 10^{-1}$  & 1611 & 1575 & 1553 \\ 
          &              & $ 3.2 - 10.0 \cdot 10^{-1}$ & 1604 & 1519 & 1464 \\ \hline
          &              &  Pointlike           &               &             &              \\ \hline
 6.6      & 3.2 - 10     & $ 3.2 - 10.0 \cdot 10^{-4}$ & 8.3 & 7.1 & 6.4 \\ 
          &              & $ 1.0 - 3.2 \cdot 10^{-3} $ & 50 & 43 & 40 \\ 
          &              & $ 3.2 - 10.0 \cdot 10^{-3}$  & 202 & 182 & 171 \\ 
          &              & $ 1.0 - 3.2 \cdot 10^{-2}$ & 644 & 602 & 579 \\ 
          &              & $ 3.2 - 10.0 \cdot 10^{-2}$  & 1846 & 1770  & 1728  \\ 
          &              & $ 1.0 - 3.2 \cdot 10^{-1}$  & 4351 & 4023  & 3846  \\ 
          &              & $ 3.2 - 10.0 \cdot 10^{-1}$  & 832 & 718 & 656 \\ \hline
 20.8     & 10 - 32      & $ 1.0 - 3.2 \cdot 10^{-3}$ & 7.0 & 6.3 & 5.9 \\ 
          &              & $ 3.2 - 10.0 \cdot 10^{-3}$  & 42 & 39  & 37  \\ 
          &              & $ 1.0 - 3.2 \cdot 10^{-2}$  & 173 & 163  & 156  \\ 
          &              & $ 3.2 - 10.0 \cdot 10^{-2}$  & 554 & 537  & 526  \\ 
          &              & $ 1.0 - 3.2 \cdot 10^{-1}$  & 1557 & 1528 & 1509 \\ 
          &              & $ 3.2 - 10.0 \cdot 10^{-1}$ & 1602 & 1518 & 1463 \\ \hline
\end{tabular}
\end{center}

\end{document}